\documentclass[10pt,journal,cspaper,compsoc]{IEEEtran}

\usepackage{graphicx}
\usepackage{times}
\usepackage{fancyhdr}
\usepackage{color}
\usepackage{hyperref}
\usepackage{amsmath}
\usepackage{amsfonts}
\usepackage{amssymb}

\usepackage{multirow}
\usepackage[latin1]{inputenc}

\newcommand{\trace}{\mathrm{Tr}}
\newcommand{\1}{{\bf 1}}

\begin{document}
%
\title{Random Walks, Markov Processes and the Multiscale Modular Organization of Complex Networks}
%
%
%
%

\author{Renaud~Lambiotte, 
        Jean-Charles~Delvenne 
        and~Mauricio~Barahona
\IEEEcompsocitemizethanks{\IEEEcompsocthanksitem Renaud Lambiotte is with Namur Center for Complex Systems (naXys) at Université de  Namur, Belgium.
E-mail: \mbox{renaud.lambiotte@unamur.be}
\IEEEcompsocthanksitem Jean-Charles Delvenne is with  Institute of Information and Communication Technologies, Electronics and Applied Mathematics (ICTEAM) and Centre for Operations Research and Econometrics (CORE) at Université catholique de Louvain, Belgium.
E-mail: \mbox{jean-charles.delvenne@uclouvain.be}
\IEEEcompsocthanksitem Mauricio Barahona is with Department of Mathematics at Imperial College London, United Kingdom. 
E-mail: \mbox{m.barahona@imperial.ac.uk}
}
\thanks{}}

%
%

%

\IEEEcompsoctitleabstractindextext{%
\begin{abstract}
Most methods proposed to uncover communities in complex networks rely on combinatorial graph properties. Usually an edge-counting quality function, such as modularity, is optimized over all partitions of the graph compared against a null random graph model.  Here we introduce a systematic dynamical framework to design and analyze a wide variety of quality functions for community detection. The quality of a partition is measured by its Markov Stability, a time-parametrized function defined in terms of the statistical properties of a Markov process taking place on the graph. The Markov process provides a dynamical sweeping across all scales in the graph, and the time scale is an intrinsic parameter that uncovers communities at different resolutions. 
This dynamic-based community detection leads to a compound optimization, which favours communities of comparable centrality (as defined by the stationary distribution), and provides a unifying framework for spectral algorithms, as well as different heuristics for community detection, including versions of modularity and Potts model. 
Our dynamic framework creates a systematic link between different stochastic dynamics and their corresponding notions of optimal communities under distinct (node and edge) centralities. We show that the Markov Stability can be computed efficiently to find multi-scale community structure in large networks.
\end{abstract}
}

\maketitle

\IEEEdisplaynotcompsoctitleabstractindextext

%
\IEEEpeerreviewmaketitle

\section{Introduction}
\IEEEPARstart{H}{ow} the structure of a network affects the 
dynamics (e.g., diffusion or synchronization)  that takes place on it
has been studied extensively in recent years~\cite{review,bocca,syn2}. 
This relationship is particularly relevant when the network is composed of tightly-knit modules or \textit{communities}~\cite{GN,simon,fortunato2010community,revporter,sales,clauset}, which
can lead, for instance, to partially coherent dynamics~\cite{syn1,neave}, or to the emergence of co-operation~\cite{cooperation} and coexistence of heterogeneous ideas in a social network~\cite{lambi}.
Conversely, it has been proposed that dynamical processes such as random walks~\cite{JC,PL,rosvall,MuchaEtAl10} and synchronization~\cite{syn1} could be used as empirical means to extract information about the network and, specifically, to uncover its community structure.

Recently, there has been extensive research on the detection of communities and hierarchies in real world systems, ranging from social systems to technological and bio-chemical systems (for a  review see~\cite{fortunato2010community}).
Most of these studies follow from the classical problem of graph partitioning and are thus based on \emph{structural} properties of graphs~\cite{fortunato2010community,revporter}. 
In order to discover communities, such methods usually proceed by optimizing a quantity that captures what is thought to be the goodness of a partition in terms of combinatorial properties of the graph. A variety of such quality functions (and associated optimization strategies) have been proposed, including different versions of balanced and normalized cuts, as well as modularity and its extensions~\cite{fortunato2010community,revporter}. In general, these combinatorial definitions operate by counting the number of links within and between the communities, and are thus blind to the flows of information taking place on the network.

In contrast, we adopt here a dynamical viewpoint for the analysis of community structure in graphs. Specifically, we use statistical properties of a random walk (or its associated Markov processes) evolving on a given network to quantify the quality of partitions \emph{across all time scales}. Consider, for instance, the simple random walk, where a random walker jumps at every step from the node where it sits to one of its immediate neighbours with a probability proportional to the weight of the link joining the nodes.  We define the 
\emph{Markov Stability}~\cite{JC,lambiotte2008laplacian,Delvenne13,schaub2012markov} 
of a partition of the graph at time $t$ as
the probability of a walker to be in the same community at time zero and at time $t$ when the system is at stationarity, discounting the expected probability as $t \to \infty$.
For an ergodic and mixing random walk (i.e., on an aperiodic, strongly connected graph), 
this limiting probability is the probability of two independent walkers to be in the same community.
The Markov Stability so defined
measures the quality of a partition in terms of the persistence of the Markov dynamics within the communities of the partition  within the time scale $t$, i.e., the Markov Stability is large when it is unlikely that  a random walker will escape the communities within time $t$. Alternatively, the Markov stability can also be understood as the time auto-correlation of a coarse-grained signal. 
Hence, a large Markov Stability is equivalent to a non-asymptotic time scale separation~\cite{SimonAndo61,kokotovic86} within the diffusion dynamics, where the fast dynamics mixes the probability flow inside the communities and the slow dynamics describes the transfer of probability between the communities.  
It can be shown that the Markov Stability so defined, which we will make more explicit below, is monotonically decreasing for most partitions on most graphs~\cite{JC}.

The dynamics-based Markov Stability framework for community detection introduced in~\cite{JC,lambiotte2008laplacian,Delvenne13} has mathematical connections with the 
wider literature relating random walks on graphs and graph properties and allows us to link those
results with applications in community detection.
A strong initial motivation for our work was the theory of quasi-stationary distributions in Markov chains~\cite{darrochseneta_discrete,darrochseneta_continuous}, and the theory of quasi-stable (long-lived) states in the physics of energy landscapes~\cite{wales2003energy}. 
Random walks have been used by a variety of methods in graph partitioning and clustering.
For example, the mixing rate of the random walker is closely related to the conductance, a measure of quality for \emph{two-way} partitions~\cite{alon1985lambda,lovasz1990mixing,chung2007four}. 
Through their commute times~\cite{YenFoussDecaestFrancqSaerens08} or through more general spectral embeddings~\cite{coifman2006diffusion}, random walks also allow representations of the graph in a Euclidean space on which classic machine learning techniques can be used, including clustering. Other partitioning algorithms have also made use of random walk measures~\cite{PL,meila2001random,piccardi2011finding,van2008graph}. The distinguishing feature of the Markov Stability approach is the systematic sweeping through all time scales, fast to slow, in order to discover fine or coarse partitions, thus relating characteristic time scales of the dynamics to the structural scales present in the network. In constrast, the precited methods focus on a fixed time scale (e.g., one-step) or a fixed number of communities (e.g., two) and hence do not exploit fully the dynamical aspects of the random walk. See~\cite{JC} for a more extensive discussion, and Section \ref{sec:discrete} for an overview of the unifying character 
of the
Markov Stability framework, whose dynamical character allows the interpolation between the 
structural (edge-counting) measures and the spectral approach to community detection.

In this article, we extend the Markov Stability formalism and show that any random walk on a given network, whether in discrete or continuous time, generates a different partition Stability function, and therefore a different notion of community reliant on specific measures of node and/or edge centrality. 
Indeed, classical notions of centrality (e.g., degree, eigencentrality, pagerank) can be shown to correspond to different random walks on the networks.
Within this framework, we observe that good communities appear as a result of an optimization that balances the cost of severing many or highly central edges 
against a maximum-entropy spread of the centrality across communities. This compound optimization is parametrically modulated by \emph{time}, which gives varying weight to the energetic cost of the \emph{cut} against the maximum entropy term. At long times, the problem turns out to be solved exactly by spectral methods.  We show how these dynamical, graph-theoretical and optimization concepts are intertwined, providing insight on the nature of different community structures, the centrality optimizations they entail, and associated spectral partitioning algorithms known in the literature.
Our work thus provides a unifying viewpoint for different variants and heuristics used in the graph-partitioning, clustering and community detection literatures, including several variants of null-model-based modularity or spectral algorithms, which appear as particular cases of our formalism.
Conceptually, our work indicates that, rather than searching for a single partition at a particular scale, dynamics can be used to unfold and detect systematically the relevant partitions by scanning across all scales in the graph~\cite{JC,Delvenne13}. Similarly, we show here that the choice of dynamics can also be used to find the most appropriate community structure (if particular information about the system is available) or to explore the network under different (and complementary) viewpoints to gain deeper information about the system.

The paper is organized as follows. First, the framework is introduced via the standard (simple) random walk and its associated continuous-time processes, including those generated by the normalized and combinatorial Laplacians. We show how the relevant centrality measure in this case is the degree, yet different continuous-time Markov processes (potentially relevant for different network dynamics) lead to different communities linked to particular heuristic null models used in the community detection literature.  The dynamical scanning implicit in our framework is used to illustrate the detection of community structure across scales in several examples without imposing the scale or number of communities \textit{a priori}.
Part of these results
were reported in the unpublished preprint~\cite{lambiotte2008laplacian}.
We then consider the analysis of less standard random walks, specifically the Ruelle-Bowen case, and show that its notion of community is based on a different kind of centrality, i.e., eigencentrality.
This is followed by a brief section where we show how the dynamical viewpoint afforded by Markov Stability seamlessly extends to the case of directed graphs, thus allowing us to recast the concept of structural communities in terms of flow communities.
The final section illustrates the framework with the analysis of synthetic benchmarks and real-world examples,  and discusses computational and practical issues for Markov Stability, e.g. assessing the presence of robust partitions, or of a hierarchical structure. 
 
\section{The simple random walk and community detection:
Discrete-time Markov Stability for undirected graphs}
\label{sec:discrete}
To make our arguments more precise, we first review briefly some of the notation and results from~\cite{JC,Delvenne13}, where mathematical proofs and further results can be found.
For simplicity, we start by considering the case of undirected graphs, although we will see below that the arguments extend to directed graphs too.

Consider an undirected graph with $N$ nodes and weighted adjacency matrix $A \in \mathbb R ^{N\times N}$, such that the weight of the link between node $i$ and node $j$ is given by $A_{ij}=A_{ji}$.
The vector containing the degrees (or strengths) of the nodes is $d = \ A \mathbf{1}$, where $\mathbf{1}$ is the $N \times 1$ vector of ones, and we also define the diagonal matrix $D = \text{diag}(d)$. The sum of all degrees is $2m = \1^T d$. The \textit{combinatorial} graph Laplacian is defined as $L=D-A$ and the \textit{normalized} graph Laplacian is defined as $\mathcal{L}=D^{-1/2} L D^{-1/2}$. Both Laplacians are symmetric nonnegative definite, with a simple zero eigenvalue when the graph is connected~\cite{chung1997spectral}. We denote the trace with the notation $\trace[ \quad ]$.

Consider the simple (unbiased) random walk governed by the standard dynamics:
\begin{equation}
\label{eq:simpleRW}
\mathbf{p}_{t+1} =\mathbf{p}_t \left[D^{-1}A \right] = \mathbf{p}_t M,
\end{equation}
where $\mathbf{p}$ denotes the $1\times N$ dimensional probability vector and $M$ is the transition matrix. Note that following the Markov chain literature, the probability vectors are defined as row vectors. Under the assumptions of a connected, undirected, and non-bipartite graph this dynamics converge to a unique stationary distribution 
\begin{align} 
\pi = d^T/2m.
\label{eq:pi}
\end{align}

Each partition of the graph into $c$ communities is encoded by a $N \times  c$ indicator matrix $H$ with $H_{ij} \in \{0, 1\}$, where a 1 denotes that node $i$ belongs to community $j$.
Given a partition $H$, the \textit{clustered autocovariance} matrix of the diffusion process at time $t$ is:
\begin{equation}
\label{eq:autocovariance}
  R_t (H) = H^T\left[\Pi M^t-\pi^T\pi\right]H ,
\end{equation}
where $\Pi = \text{diag}(\pi)$. 
The $c \times c$ matrix $R(t)$ reflects the probability of the random walk to remain within each block (diagonal elements) and to transfer between blocks (off diagonal elements) after a time $t$. 
Consequently, we define the Markov Stability of the partition $H$ as
\begin{equation}
\label{eq:stability_disc}
r_t (H) = \min_{0\leq s\leq t}\trace\left[R_s(H)\right] \approx \trace \left[R_t(H)\right],
\end{equation}
the approximation coming from the computational observation that $\trace \left[R_t(H)\right]$ is mostly monotonically decreasing for empirical graphs~\cite{LeMartelot2012}.
A `good' partition over a time scale $t$ has well-defined communities that preserve probability flows within them, hence maximizing the trace of $R_t$ and, conversely,  
the Markov Stability $r_t(H)$ can be seen as a quality function for a partition of a graph as a function of the time horizon of the random walk.
\begin{figure}[!ht]
\begin{center}
\includegraphics[width=0.5\textwidth]{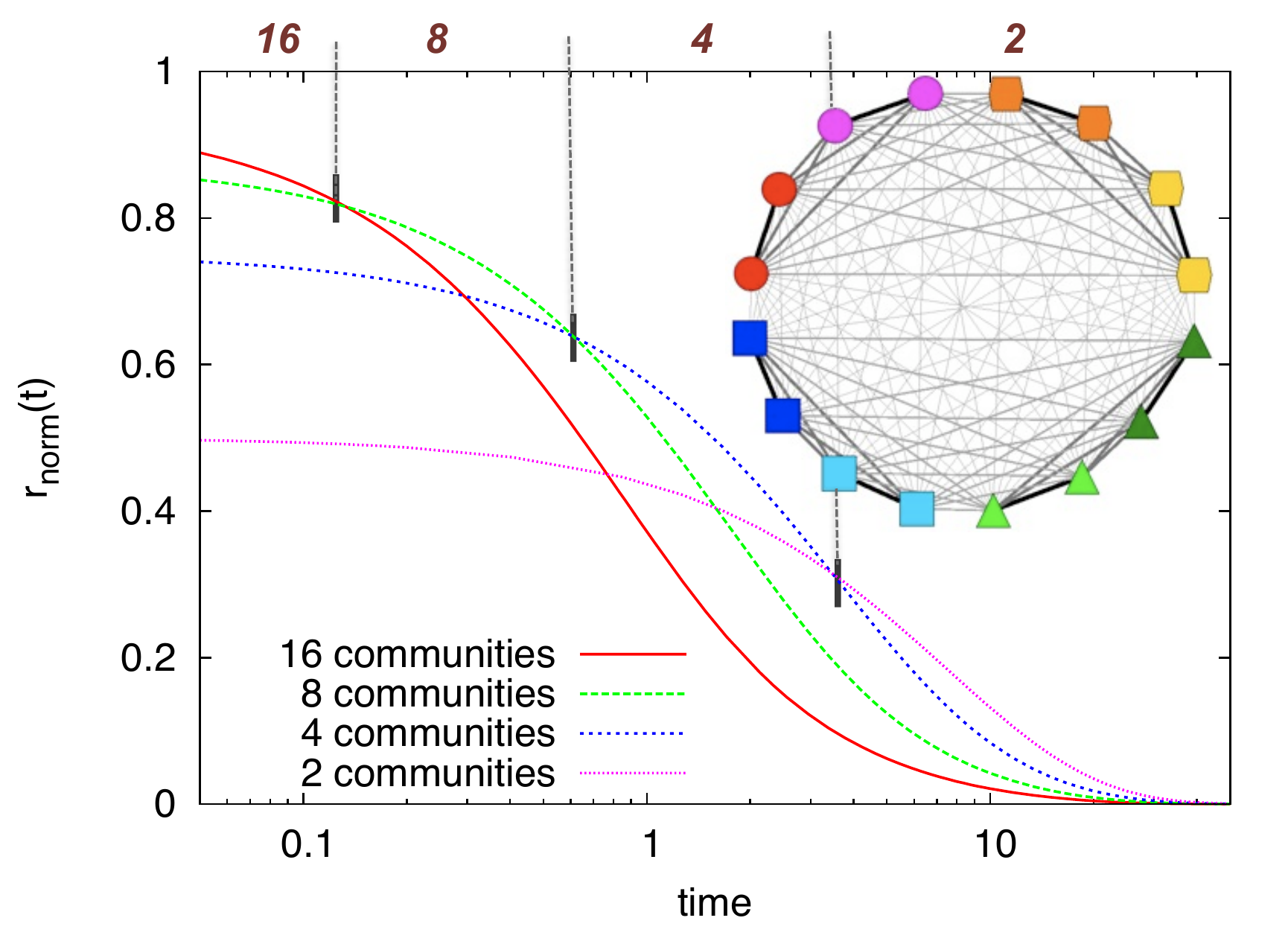}
\end{center}
\caption{{\bf Unfolding the multiscale community structure of a hierarchical network 
as a function of Markov time.}
\small{As an illustration, consider a hierarchical graph generated as follows~\cite{hier}: start with a pair of nodes connected by a link of weight $c<1$, duplicate them and add a
link of weight $c^2$ between all pairs of nodes in different modules. 
Iterate the procedure $K$ times to obtain a fully connected, weighted network of $2^K$ nodes.  The figure shows a network with $2^4=16$ nodes with edges shaded according to their strength ($c=1/4$).
By symmetry, the natural partitions are into $16$ single nodes, $8$ pairs (colours), $4$ tetrads (shapes) and $2$ groups of $8$ nodes (upper and lower hemispheres). Evaluation of the Markov Stability $r_\text{norm}(t)$ shows that, as $t$ grows, the optimal partition goes from 16 communities to 8 to 4 to 2 over different time intervals.}
}
\label{fig2}
\end{figure}

The Markov Stability $r_t(H)$ can be used to rank partitions of a given graph at different time scales or,  alternatively,  $r_t(H)$ can be used as an objective function to be maximized for every time $t$ in the space of all possible partitions of the graph:
\begin{equation}
\label{eq:stability}
r_t = \max_{H} r_t(H).
\end{equation}
Such an optimization results in a sequence of partitions optimal over different time interval. Although this optimization is NP-hard, a variety of efficient optimization heuristics for graph clustering can be used, as discussed in later sections. 

The discrete-time Markov Stability $r_t(H)$ for undirected graphs encompasses several 
well-known heuristics and has other desirable theoretical properties, some of which we highlight here succinctly (see~\cite{JC,Delvenne13} for proofs):
\begin{itemize}
\item Discrete-time Markov Stability at time $t=1$ is equal to 
the `usual' modularity $Q_{\mathrm{conf}}$, i.e., 
with the \emph{configuration model} as null model~\cite{newman_modul_PNAS,GN}: 
\begin{equation}
\label{eq:modul_stab}
r_1(H) = \trace\left[ H^T\left(\dfrac{A}{2m} -\pi^T\pi\right)H\right] = Q_{\mathrm{conf}}.
\end{equation}
\item Markov Stability at time $t=0$ is equivalent to the Gini-Simpson diversity index of the partition $H$~\cite{simpson1949measurement}:
\begin{equation}
\label{eq:divers}
 r_0(H)  
 = 1-\sum_{\mathcal{C}=1}^c (\pi h_\mathcal{C})^2 
 = \text{GS}_{\pi},
\end{equation}
where $h_\mathcal{C}$ is the $\mathcal{C}$-th column of the matrix $H$.
GS$_{\pi}$ is a measure of entropy of the partition according to the values of $\pi$, i.e.,
the degree. 
GS$_{\pi}$ is large when the partition has many communities of equal size (according to $\pi$), and is low when the partition has few and uneven communities. GS$_{\pi}$ is maximum for the partition into one-node communities.
This index is well known in economics (Hirschman-Herfindahl index~\cite{hirschman1964paternity}) and information theory (R\'enyi entropy~\cite{renyi1961measures}), among others.
\item The probability of changing community in one step
\begin{align}
\label{eq:cut}
r_0 (H)-r_1(H)=1-\trace\left[H^T\frac{A}{2m} H\right] = \text{Cut},
\end{align} 
is a measure of the cut induced by the partition, i.e., 
the fraction of edges between all the communities. 

\item The long-term behavior of $r_t$ is governed by the normalized Fiedler eigenvector associated with the second dominant eigenvalue of $M$, i.e., that which is closest to $1$ in absolute value. Hence the optimal community structure as $t\to\infty$ is typically~\footnote{Close-to-bipartite graphs are the exception: they have a strongly negative eigenvalue whose odd and even powers generate an alternating $r_t$.} 
given by the bipartition according to the sign of the entries of the normalized Fiedler eigenvector~\cite{JC,Delvenne13}. 

\item  Spectral algorithms (either iterative or based on several eigenvectors at a time) are classic relaxation heuristics~\cite{Fiedler73,Fiedler75} for the optimization of a variety of NP-hard partitioning quality functions, including modularity~\cite{Newman06} or normalized cut~\cite{ShiMalik00}. We have shown that spectral clustering methods provide \emph{exact} procedures for the optimization of Markov Stability at long times.

\end{itemize}

\section{Continuous-time Markov Stability: the dynamical origin of different quality functions}

We now consider continuous-time Markov processes associated with the \emph{simple} random walk~\eqref{eq:simpleRW} in order to extend our dynamics-based framework for community detection in \emph{undirected} graphs. 

\subsection{Normalized Laplacian Markov Stability}
Given the random walk~\eqref{eq:simpleRW} on an undirected graph, a standard way to derive a continuous-time model is to assign a continuous Poisson process of given density at each node~\cite{Bremaud,tijms2003first}.  
If we assume identically distributed Poisson processes (i.e., with identical waiting times) for all nodes, we obtain the standard diffusive dynamics:
\begin{equation}
\label{eq:norm}
\frac{d\mathbf{p}}{dt} = - \mathbf{p}\;[I - D^{-1} A] =  - \mathbf{p}\;[D^{-1}L]
\end{equation}
Note that the operator $D^{-1}L$ is isospectral with the normalized Laplacian $\mathcal{L}$ since they are related by the similarity transformation
$D^{-1/2} \mathcal{L}  D^{1/2} = D^{-1}L$. Hence the dynamics of~\eqref{eq:norm} is dictated by the spectral properties of $\mathcal{L}$. In particular, this process
converges to the same unique stationary distribution~\eqref{eq:pi} as the (discrete-time) simple random walk. 
As above, we thus define the \emph{continuous-time Markov Stability} as:
\begin{align}
\label{eq:stability_cont}
r_{\text{norm}}(t; H) =  
\trace \left[ H^T\left(\Pi e^{-t\,D^{-1}L} -\pi^T\pi\right)H \right], 
\end{align}
where the notation $r_{\text{norm}}$ emphasizes the connection with the normalized Laplacian. This continuous-time version of Markov Stability shares broadly similar properties with the discrete-time version~\eqref{eq:stability_disc}, and most of the discussion presented in Section~\ref{sec:discrete} applies here.  For instance, Figure~\ref{fig2} shows the results of the optimization of 
$r_{\text{norm}}(t; H)$ over time and over the space of partitions for a simple example. Note that the Markov Stability explores the community structure at all scales (from finer to coarser) using the dynamic zooming provided by the Markov time of the diffusion process $t$. The relevant (time) scales emerge as the ones  leading to persistent (robust) partitions over extended intervals of time.  See Section~\ref{sec:methods}~and~Refs.~\cite{Delvenne13,schaub2012markov} for a discussion of some of the practical issues of the computational implementation and more illustrative examples.

It is also instructive to consider the behavior of~\eqref{eq:stability_cont}
in the limit of small times,  $t \to 0$. Keeping terms to first order, we obtain the \emph{linearized Markov stability}:
\begin{align}
r_\text{norm}^\text{lin}(t;H) 
&= r_\text{norm}(0;H) - t \,\, \trace\left[H^T \frac{L}{2m} H\right]  \nonumber \\
&= \text{GS}_{\pi} - t \, \text{Cut} \label{eq:norm_lin_cut} \\
& = (1-t) \, \text{GS}_{\pi} + t \, Q_\text{conf} \label{eq:norm_lin_potts}
\end{align}
where we have used~\eqref{eq:modul_stab}--
\eqref{eq:cut} and the fact that $\trace\left[H^T L H\right ] =  2m - \trace\left[H^TAH\right] $. 
A few remarks about the linearized Markov Stability follow:
\begin{itemize}
\item Analogously to~\eqref{eq:cut}, the instantaneous probability rate of the walker escaping from its initial community $- \left. d r_\text{norm}(t;H)/dt \right |_{t=0}= \trace [H^T L H]/2m $ 
is the Cut.

\item The Potts model heuristic proposed by Reichardt \& Bornholdt~\cite{ReichardtBornholdt06} is exactly recovered as the linearized Markov stability. Hence we can see the Markov time $t$ as the equivalent of a resolution parameter. From~\eqref{eq:norm_lin_potts} it also follows that the `usual' modularity~\cite{newman_modul_PNAS,GN} is recovered at $t=1$ for undirected graphs:
\begin{align}
r_\text{norm}^\text{lin}(1;H) =  Q_\text{conf}.
\end{align}

\item Equation~\eqref{eq:norm_lin_cut} provides an interpretation of Markov Stability as a compound quality function to be optimized under two competing objectives: minimize the Cut size while trying to maximize the diversity $\text{GS}_{\pi}$, which favours a large number of equally-sized communities according to $\pi$, thus resulting in more balanced partitions.  The relative weight between both objectives is modulated as the Markov time $t$ increases. 
\end{itemize}

The stationary distribution $\pi$ plays a key role in the 
definition of the community quality function: 
\begin{itemize} 
\item Firstly, $\pi$ can be understood as originating the \emph{null model} of modularity, i.e., the model of random graph against which the network is compared to detect the significance of the communities. The null model in the `usual' modularity is the configuration model, which randomly rewires the edges of a given graph preserving the degree of every node. The probabilistic description of this model is given by the outer product $\pi^T \pi$, which in our dynamical interpretation corresponds to the expected transfer probabilities at stationarity for this Markov process. 
\item Secondly, $\text{GS}_{\pi}$ measures the diversity of the partitions according to the node property $\pi$. Hence, as the value of $t$ grows, the optimization leads to balanced distributions of $\pi$ across communities, splitting nodes with high values of $\pi_i$ into different communities. In this case, we tend to segregate nodes with high degree into different groups. 
\end{itemize}

\subsection{Combinatorial Laplacian Markov Stability}

Given a discrete-time random walk, a variety of continuous-time Markov processes are possible. 
Although in~\eqref{eq:norm} we assumed identical Poisson processes at all nodes, we have the flexibility to assign different waiting times at each node.
An interesting choice is to consider that the waiting time at each node is inversely proportional to its degree, i.e., the walker spends less time on nodes of high degree.
Using an inhomogeneous rescaling of time this leads to a Markov process governed by the \emph{combinatorial Laplacian}:
\begin{align}
\label{eq:comb}
\frac{d\mathbf{p}}{dt} D^{-1} \langle d \rangle &= - \mathbf{p} \, D^{-1} \; [D - A]  \nonumber \\
 \implies \frac{d \mathfrak{p}}{dt}  &=   -\frac{1}{\langle d \rangle} \mathfrak{p} \; L,
\end{align}
where $\langle d \rangle = (\1^T D \1)/N$ is the average degree and $\mathfrak{p}=~\mathbf{p} D^{-1}$. 
The stationary distribution of~\eqref{eq:comb} is now the uniform distribution over the nodes: 
\begin{align}
\pi_c = \1^T/N, 
\end{align}
and the \emph{combinatorial} continuous-time Markov Stability is:
\begin{align}
\label{eq:stability_cont_comb}
r_{\text{comb}}(t; H) =  
\trace\left[ H^T\left(\Pi_c e^{-t\,L/{\langle d \rangle}} -{\pi_c}^T\pi_c\right)H \right]. 
\end{align}
The corresponding linearized version is then:
\begin{align}
r_\text{comb}^\text{lin}(t;H) 
&= \text{GS}_{\pi_c} - t \, \text{Cut} \label{eq:comb_lin_cut} \\
& = (1-t) \, \text{GS}_{\pi_c} + t \, Q_\text{ER}. \label{eq:comb_lin_potts}
\end{align}

In this case, the stationary distribution $\pi_c$ leads to a different diversity index:
\begin{align}
\label{eq:divers_comb}
\text{GS}_{\pi_c} 
 =  1 - \sum_{\mathcal{C}=1}^c (\1^T h_\mathcal{C}/N)^2 
 = 1 - \sum_{\mathcal{C}=1}^c \left (n_\mathcal{C}/N \right)^2,
 \end{align}
 where $n_\mathcal{C}$ is the number of nodes of community $\mathcal{C}$.
The modularity associated with this process is:
 \begin{align}
Q_\text{ER} 
= \text{GS}_{\pi_c} - \text{Cut}=
\trace \left[ H^T \left(\frac{A}{2m}- \frac{\1 \1^T}{N^2} \right) H \right],
\end{align}
which is precisely the modularity based on the Erd\"os-R\'enyi (ER) null model with a probabilistic description given by the outer product $\1 \1^T/N^2$. 
This version of modularity was originally discussed by Newman~\cite{GN,newman_modul_PNAS} and has been recently studied against network benchmarks~\cite{traag2011narrow}.  
Based on our arguments above, the combinatorial Markov Stability optimizes partitions that balance the Cut against the diversity $\pi_c$, which ignores degrees and counts only the fraction of nodes present in each community. Hence, it is more likely to group nodes with high degree in the same community when using combinatorial Markov Stability, as we will discuss below.

Finally, we remark that at long Markov time scales, the combinatorial Laplacian dynamics recovers the bipartition based on the classic heuristic of the signs of the components of the Fiedler eigenvector~\cite{Fiedler73}, which constitutes the basis of several spectral algorithms. As stated above, the normalized Laplacian version converges to the bipartition based on the normalized Fiedler eigenvector, which is also used in other spectral algorithms like Shi-Malik~\cite{ShiMalik00}. Seeing those algorithms as the coarser extreme of a range of community detection problems provides additional insight into the meaning and differences between those popular spectral algorithms.

\subsection{Normalized vs Combinatorial Markov Stability: some examples}

\subsubsection*{The relevance of dynamical coherence}

As discussed above, a driving force in the definition of quality functions for community detection has been the use of null models, i.e., random graph models that preserve certain properties of the graph under study and act as bootstraps to establish the significance of communities.  
Early on, it was proposed~\cite{GN,newman_modul_PNAS} that the configuration model should be preferable to Erd\"os-R\'enyi as the null model, because the former takes into account the degree heterogeneity typically found in realistic networks.   
However, it has been recently shown\cite{traag2011narrow} that the Erd\"os-R\'enyi  model behaves at least as well as the configuration modularity on benchmarks~\cite{LancichinettiEtAl08} and leads to improved results in particular graphs.

Under our dynamical framework, the two null models correspond to the stationary distributions of the Markov processes governed by the normalized and combinatorial Laplacians. 
The two Laplacian dynamics can emerge naturally in the modelling of different continuous-time dynamics on networks, such as heat diffusion~\cite{kondor2002diffusion, chung1997spectral}, the linearization of Kuramoto oscillators~\cite{syn1,syn2}, or consensus dynamics 
~\cite{ali,ROS-JAF-RMM:07,neave}. 
In the important cases when the dynamics of the system is governed by the combinatorial Laplacian (e.g., synchronization, consensus, or vibrational dynamics), we expect that the relevant dynamical groupings should correspond to communities obtained using the combinatorial version of Markov Stability (i.e., corresponding to the ER null model) and not the canonical configuration model.

\begin{figure}[t]
\begin{center}
\includegraphics[width=0.5\textwidth]{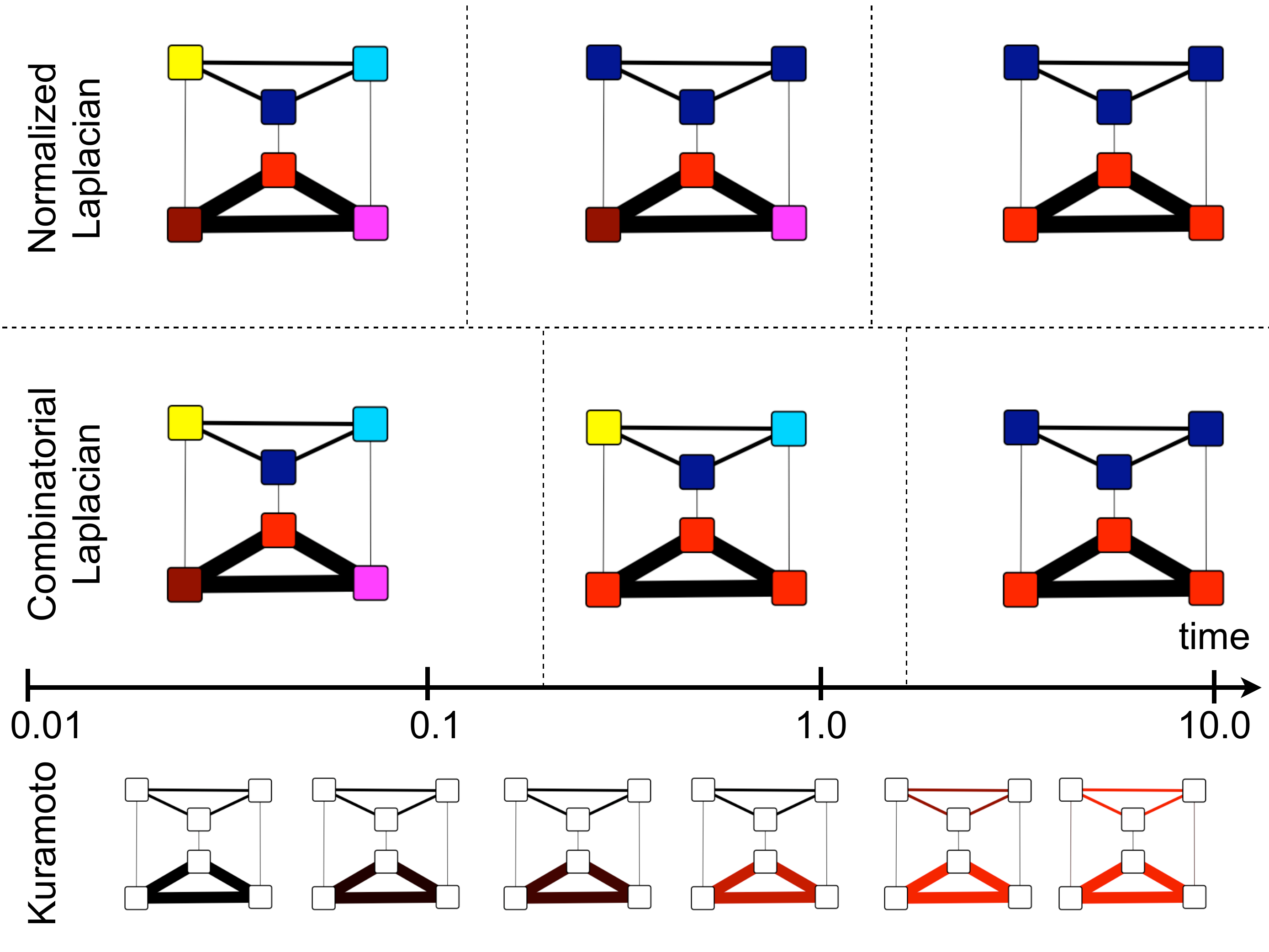}
\end{center}
\caption{{\bf Dynamical coherence in synchronization and community structure.} 
\small{We computed the coherence of Kuramoto oscillators in this toy network and represented it 
in the bottom panel by using a colour code, from black to red as the coherence grows. 
The lower triangle is always more coherent than the upper triangle. 
The partitions obtained by optimizing the combinatorial Markov Stability $r_{\text{comb}}(t; H)$, related to the Erd\"os-R\'enyi null model, capture this behavior. On the other hand, the optimization of the  normalized Markov Stability $r_{\text{norm}}(t; H)$, related to the usual configuration model, does not find the relevant sequence of partitions. 
}}
\label{fig4}
\end{figure}

Figure~\ref{fig4} illustrates this point by examining relevance of dynamic communities in synchronization dynamics on a toy network made of two triangles: 
links in the upper triangle have weight 5; links in the lower triangle have weight 25; and they are connected by links of weight 1. 
The dynamics of the network is given by the Kuramoto model with uniform frequencies, 
a prototypal model for synchronization where each node has a phase $\phi_i$ evolving as 
\begin{equation}
\dot{\phi}_{i} = \omega + \sum_{j} A_{ij} \sin(\phi_j-\phi_i).
\label{eq:kuram}
\end{equation}
The coherence between nodes $i$ and $j$ is measured by the order parameter $\rho_{ij}(t)= \langle \cos\left(\phi_i(t)-\phi_j(t)\right) \rangle_{IC}$, where the average is performed over an ensemble of random initial conditions.  
The coherence $\rho_{ij}(t)$ computed from simulations (bottom panel)
shows that the lower triangle is always more coherent than the upper triangle, as expected. 
If we threshold to find coherent clusters~\cite{syn1}, the first group detected is the lower triangle, followed by the upper triangle at later times. 
If we use the combinatorial Markov Stability on this toy graph, this sequence of partitions is correctly uncovered. This follows unsurprisingly from our dynamical interpretation since the linearization of the Kuramoto dynamics leads to the combinatorial Laplacian. In contrast, $r_\text{norm}(t)$ does not recover this result, as it first uncovers a dynamically irrelevant partition where the upper triangle is found.
Interestingly, numerics on Kuramoto dynamics~\cite{syn1, arenas2} have shown
that the `usual' modularity $Q_{\rm conf}$ is only optimized for near-regular graphs, i.e., when it is equivalent to the true optimization performed by the dynamics, $Q_{\rm ER}$.  
Therefore, if we are interested in coherent Kuramoto communities (e.g., motivated by power grid applications~\cite{dorfler2013synchronization}),  
the partitions found with the `usual' modularity could be misleading. 
On the other hand, if we are interested in the study of probabilistic diffusive dynamics, the relevant communities should follow from the study of  $r_\text{norm}(t)$.

\begin{figure}[!ht]
\begin{center}
\includegraphics[width=.5\textwidth]{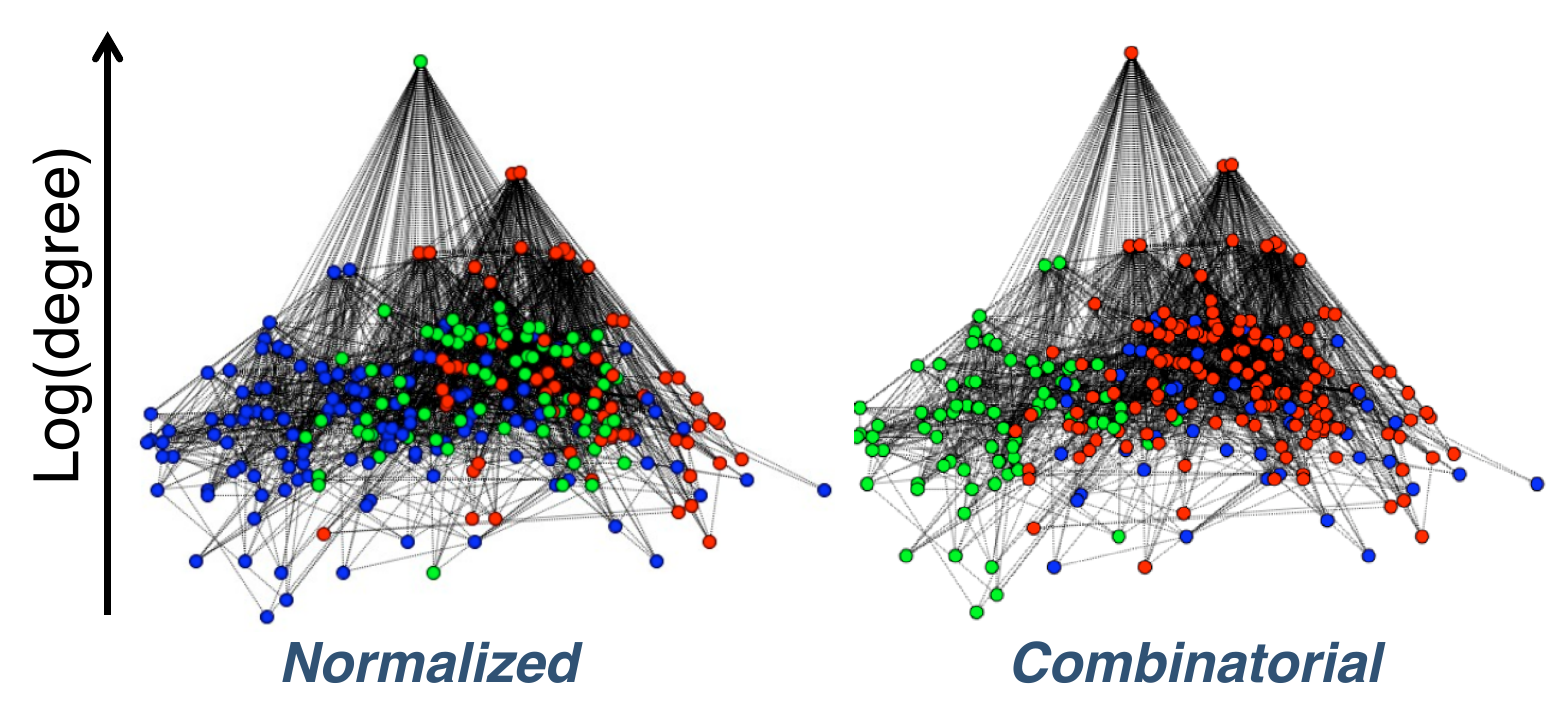} 
\end{center}
\caption{{\bf Different random walks, different community structure: the 
C. Elegans neural network.}
\small{The choice of Laplacian dynamics leads to different communities in this real-life
example. Here we present the partitions at $t=7.8$ that optimize $r_\text{norm}$ (left) and $r_\text{comb}$ (right) consisting mainly of 3 large communities in both cases (indicated by different colors). 
The nodes are displayed along the vertical axis according to their degree centrality.
The normalized Laplacian Markov Stability biases towards equicentral communities thus leading to a separation of high degree nodes into different communities, whereas high degree nodes can be grouped within the same community for the combinatorial Laplacian version.  
}  
}
\label{fig5New}
\end{figure}

\subsubsection*{An optimization perspective: distinct cost functions}

Further insight into the communities for each version of Markov Stability can be gained by
examining the role of the stationary distribution of the Markov process in the definition of the diversity index appearing in the compound cost function to be optimized. 
From the definitions~\eqref{eq:divers}~and~\eqref{eq:divers_comb} of the diversity indices GS$_\pi$ and GS$_{\pi_c}$ (associated with the normalized and combinatorial versions of Markov Stability, respectively), it follows that the normalized version balances communities with respect to their edge volume while the combinatorial version balances communities with respect to their node volume. Therefore, the normalized version (related to the `usual' modularity) tends to separate nodes with high degree into different communities. This may lead to unexpected results, e.g., in assortative networks, where high degree nodes tend to be strongly connected to one another, yet could be split when using quality functions based on the configuration model.

To illustrate this point, consider the community structure uncovered in the symmetrized version of the C. elegans neural network, a weighted network with $297$ nodes and $2m=17598$ edges. 
The partitions found by the combinatorial and normalized versions of Markov Stability are significantly different---not unexpectedly since the graph is far from being degree-homogeneous.
In Fig.~\ref{fig5New}, we present the partitions at $t=7.8$ for both versions
consisting of mainly 3 large communities.
As discussed, the optimization of $r_\text{norm}(t)$ tends to balance the total degree $\sum_{i \in \mathcal{C}} d_i$ of the communities $C$, while  $r_\text{comb}(t)$ tends to balance the number of nodes $n_\mathcal{C}$ of the communities. 
Indeed, for the combinatorial Laplacian, the total degree of each of the three communities are $\{1984, 11782, 3424 \}$, whereas these numbers are more balanced for the normalized Laplacian:  $\{5753, 5561, 6284 \}$.
On the other hand, the fact that the combinatorial Markov Stability does not penalize as much grouping together nodes with high degree into the same community can also be seen in Fig.~\ref{fig5New}.
The high degree nodes tend to be split evenly among the three communities for the normalized Laplacian, while the combinatorial Laplacian has a disproportionately large number of high degree nodes grouped together in the red community, less so in the green community and even fewer in the blue community. More specifically, the top 20 nodes with the highest degree are distributed among the three communities in the ratios $\{18, 2, 0\}$ for  $r_\text{comb}$ while the corresponding ratios for $r_\text{norm}$ are  $\{13, 5, 2\}$.

\subsection{The simple random walk and its continuous-time versions: degree as centrality}

Our discussion above leads to the following generalization of the continuous-time versions of the simple (unbiased) random walk.  When taking the continuum limit, the waiting times at each node can be weighted by any power of the degree: 
\begin{align}
\label{eq:general}
\frac{d\mathbf{p}}{dt} D^{k} \langle d^{-k} \rangle &= - \mathbf{p} \, D^{k} \; D^{-k} [I - D^{-1}A]  \nonumber \\
 \implies 
 \frac{d \mathfrak{p}}{dt}  &=   -\frac{1}{\langle d^{-k} \rangle} \mathfrak{p} \; L_k,
\end{align}
where the notation $\langle \ldots \rangle$ denotes the average over all the nodes, i.e., $\langle d^{-k} \rangle = (\1^T D^{-k} \1)/N$, 
and we have introduced the $k$-scaled Laplacian:
\begin{align}
\label{eq:k-Laplacian} 
L_k= D^{-k} [I - D^{-1}A].
\end{align}
The stationary distribution of~\eqref{eq:general} is then
\begin{align}
\label{eq:k-stationary}
\pi_k =\1^T D^{k+1}/ \left(\1^T D^{k+1} \1\right), 
\end{align}
and the corresponding $k$-scaled Markov Stability is:
\begin{align}
\label{eq:stability_cont_k}
r_{k}(t; H) =  
\trace\left[ H^T\left(\Pi_k e^{-t\,L_k/{\langle d^{-k} \rangle}} -
{\pi_k}^T\pi_k\right)H \right]. 
\end{align}
The linearized version reads:
\begin{align}
r_k^\text{lin}(t;H) 
&= r_k(0;H)
 - t \, \trace \left[ H^T \left (\frac{L}{N \langle d^{k+1} \rangle \langle d^{-k} \rangle} \right) H \right] \nonumber \\
&= \text{GS}_{\pi_k} - t \, \frac{\langle d \rangle}{\langle d^{k+1} \rangle \, \langle d^{-k} \rangle} \, \text{Cut}, \label{eq:k_lin_cut}
\end{align}
and, again, the diversity index of the partition is measured as a function of the stationary distribution $\pi_k$: 
\begin{align}
\label{eq:divers_k}
\text{GS}_{\pi_k} 
 =  1 - \sum_{\mathcal{C}=1}^c \left(\1^T  D^{k+1} h_\mathcal{C}/ 
 \left(\1^T D^{k+1} \1\right) \right)^2. 
\end{align}
Clearly, $k=0$ corresponds to making the waiting time independent of the degree and leads to the normalized Laplacian Markov Stability, while $k=-1$ corresponds to making the waiting time inversely proportional to the degree and produces the combinatorial Laplacian version.

This generalization allows us the flexibility to modulate the effect of degree centrality in community detection using other continuous-time dynamics. We could consider a model where the waiting time is proportional to the degree, i.e., $k=1$. This could be interpreted as the model of a random web surfer, spending on average more time reading a page with higher number of links. The community detection on such a system would then be based on the non-standard Laplacian $L_1=D^{-1}-D^{-2} A$ and the diversity index~\eqref{eq:divers_k} will try and balance communities according to the square of the degree, making it even more unlikely to group high degree nodes in the same community.
If, on the contrary, we consider a model where the waiting times have an inverse square dependence on the degree ($k=-2$), the diversity index~\eqref{eq:divers_k} would then be based on the inverse of the degree, and the community detection will tend to push neighboring high degree nodes together in a single community, while low degree nodes stand separated, as in a core-periphery decomposition.  
This phenomenon will be more acute as we make $k$ more negative, whereas, 
conversely, a large and positive $k$ will put the emphasis on separating the few top degree nodes, disregarding almost entirely the effect of the majority of nodes.

This extended discussion of the simple random walk and associated Markov processes highlights
the connection of dynamical community detection with concepts of centrality. 
Measures of centrality aim at rating how connected nodes are with the rest of the network. The weighted degree is perhaps the most elementary concept of centrality---indeed, it is sometimes referred as `degree centrality'.  As shown above, the degree appears as the stationary distribution of the simple random walk~\eqref{eq:simpleRW}, and the optimization of the quality function for community detection balances the partitions according to the diversity of degree centrality.  In particular, it is optimal to split apart highly central nodes (i.e., with high degree in this case) into different communities for short enough Markov time scales, and to aim towards balanced intra-community edge centrality.  The continuous-time versions are able to modulate, amplify, attenuate, cancel or even invert the effect of degree centrality as the power $k$ is varied.  We consider the connection of dynamical community detection with other measures of centrality in the following section.

\section{Community detection based on other notions of centrality: the Ruelle-Bowen random walk}

\subsection{The role of centrality in community detection}
In different applications, it might be desirable to employ other measures of centrality as the linchpin for community detection. We can achieve this using the random-walk framework discussed above.  Many discrete-time random walks other than the simple random walk may be performed on a network.  We then may think of the stationary distribution of every random walk as a \textit{centrality measure}. Every random walk with transition matrix $M$ will then be associated with a dynamical Markov Stability quality function, and the corresponding community detection will produce optimized partitions which are balanced according to different measures of centrality.  A generic way to generate random walks is to bias the simple random walk~\cite{lambiotte2011flow}. For instance, one may attribute a positive number $b_i$ to every node $i$ (e.g., a property related to a measure of centrality) and let a random walker at $i$ jump to $j$ with probability proportional to $b_i A_{ij} b_j$.  

Once the discrete-time random walk (and its associated centrality) is chosen, different continuous-time processes can be obtained. Generically, this is done by combining two ingredients: the transition probabilities of the discrete-time random walk (i.e., the row-stochastic matrix $M$) and the waiting times of the continuous-time process at each node (compiled in a node vector $w$). The resulting process is then:
\begin{equation}
\frac{d\mathbf{p}}{dt}=-\mathbf{p}\;W^{-1}(I-M)
\label{eq:generalCTMC}
\end{equation}
with $W=\text{diag}(w)$.
These two ingredients come into play differently in determining the corresponding Markov Stability function for community detection.
The \emph{discrete-time} random walk defined by $M$ determines the 
stationary distribution $\pi_\text{disc}$ on nodes. 
On the other hand, the \emph{continuous-time} stationary distribution on node $i$, or node centrality, is given by $w_i \pi_{\text{disc},i}/\langle w \rangle$, where  $\langle w \rangle$ is the normalization constant $\pi_\text{disc} w$. As shown in the examples above, the choice of waiting times can thus modulate the effect of the node centralities. 
The centrality of \emph{edge} $ij$, on the other hand, is the probability that an observed transition links $i$ to $j$, which does not depend on the time elapsed between transition but rather on the respective frequencies of transitions given by $\pi_{\text{disc},i} M_{ij}$. Edge centralities are therefore given by $\Pi_\text{disc}M$, hence completely determined by the discrete-time transitions and unaffected by waiting times. 
As a result, the discrete-time transitions and waiting times have a different effect on the resulting Markov Stability function: waiting times have no influence on the edge centrality but afford complete control over the node centrality (and on the Gini-Simpson term of the cost function), whereas the Cut term is completely determined by the edge centralities (i.e., the underlying discrete-time random walk). At long times, the optimal split is provided by the sign pattern of the second eigenvector of the `generalized Laplacian' $W^{-1}(I-M)$, which depends both on the discrete-time transitions $M$ and the waiting times $W$.
We now explore a classic discrete-time random walk with distinctive properties. 

\subsection{Community detection according to the Ruelle-Bowen random walk}
A particularly interesting example is the random walk introduced by Ruelle, Bowen and others~\cite{Ruelle1978}. Consider a graph with adjacency matrix $A=A^T$, under the usual assumptions of connected, undirected, and non-bipartite, for simplicity.
An important notion of centrality is associated with $\mathbf{v}$, the dominant eigenvector of A (i.e., the eigenvector with the largest eigenvalue):
\begin{align}
A \mathbf{v} = \lambda_1 \mathbf{v}.
\end{align} 
The eigencentrality~\cite{Bonacich72} of node $i$ is given by $v_i$, its correspondent component of this eigenvector.  

The discrete-time Ruelle-Bowen (RB) random walk is defined such that the transition between nodes $i$ and $j$ occurs with probability $v_i A_{ij} v_j$:
\begin{equation}
\label{eq:RB-RW}
\mathbf{p}_{t+1} =\mathbf{p}_t \left[ \frac{1}{\lambda_1}  \Delta_{\mathbf{v}}^{-1}A \Delta_{\mathbf{v}} \right] = \mathbf{p}_t \, M_\text{RB},
\end{equation}
with $\mathbf{p}$ the $1\times N$ probability vector and
$\Delta_{\mathbf{v}} = \text{diag}(\mathbf{v})$. 
Under such assumptions, the unique stationary distribution of the RB random walk is 
\begin{align}
\label{eq:RB-stationary}
\pi_\text{RB} =\1^T \Delta_{\mathbf{v}}^2/ \left(\1^T \Delta_{\mathbf{v}}^2 \1\right) =
\1^T \Delta_\mathbf{v}^2, 
\end{align}
since $\left(\1^T \Delta_{\mathbf{v}}^2 \1\right) = \mathbf{v}^T \mathbf{v} =1$ for the normalized eigenvector.
The stationary distribution $\pi_\text{RB}$ can be seen as a centrality measure, which is called entropy rank (for the unweighted case) or free energy rank (for the weighted case)~\cite{delvenne2011centrality}, thus essentially equivalent to eigencentrality in terms of ranking (although the concepts diverge in the directed case, not analyzed here).

This classic random walk has an interesting interpretation in terms of entropy: it is maximally exploratory in the sense that its per-step entropy is maximal. More precisely,  let $h$ denote the (Kolmogorov-Sinai) entropy rate of the random walk, which is the average per-step entropy that is asymptotically approached for long paths, and let $E$ be the expectation of the edge transition energies $E_{ij}$, such that $A_{ij}=\exp(E_{ij})$. 
Then the RB random walk maximizes the `free energy' $h+E$. It therefore tends to make all paths of same length equiprobable, with a bias to make high energy paths more probable
~\cite{parry1964intrinsic}.
Beyond its thermodynamic properties, the Ruelle-Bowen walk naturally emerges in other contexts, such as the computation of quasi-stationary distributions~\cite{darrochseneta_discrete,darrochseneta_continuous}.

Similarly to the simple random walk, we can associate continuous-time Markov processes to the RB random walk. The simplest is given by the homogeneous waiting times:
\begin{align}
\label{eq:RB-continuous}
 \frac{d \mathbf{p}}{dt}  &=  - \mathbf{p} \;  [I - M_\text{RB}] 
\end{align}
with $M_\text{RB}$ as in~\eqref{eq:RB-RW}.
The node stationary distribution of~\eqref{eq:RB-continuous} is given by~\eqref{eq:RB-stationary},
whereas the edge centralities are given by the matrix $\Delta_{\mathbf{v}}^2 M_{\text{RB}}=\Delta_{\mathbf{v}} A \Delta_{\mathbf{v}}/\lambda_1$.
The full and linearized versions of the RB Markov Stability follow closely the expressions in~\eqref{eq:stability_cont}--\eqref{eq:norm_lin_potts}.
This continuous-time process can be generalized through the choice of waiting times.

The RB Markov Stability has connections with other heuristics in the
literature. For instance, the spectral algorithm associated with the RB random walk
on an undirected graph makes use of the second eigenvector of the adjacency matrix $A$, 
similarly to the `adjacency spectral clustering' of Sussman \textit{et al.}~\cite{sussman2012consistent}. 
To illustrate the flexibility of the framework in designing cost functions associated to different notions of communities, let us consider waiting times $W=D \Delta_v^{-2}$. This choice makes thenode centralities proportional to the degree, since the discrete-time RB walk induces stationary probability on nodes proportional to $\Delta_v^{2}$ (see Eq. (\ref{eq:RB-stationary}), while the edge centralities, unaffected by waiting times, are still determined by the edge entropy rank. The linearized Markov Stability optimization will now look for communities balanced in terms of number of edges (through diversity term) while cutting edges with low entropy rank (through the Cut term).

As a simple example of the impact of such a choice on the outcome of partitioning, consider the graph $A-B-C$ composed of two $N$-cliques $A$ and $B$ and a $N$-cycle $C$, interconnected by single edges. From the point of view of the simple random walk Markov Stability, cutting the $A-B$ edge or the $B-C$ edge is indifferent as far as the cut term is concerned. However, RB Markov Stability favours cutting the less central $B-C$ edge, thus isolating first the `hollow' module $C$ on the account of cut minimization, while the  Gini-Simpson term tends in this case to keep apart high-degree nodes, thus inducing non-trivial results~\cite{ochab2013maximal}. 
This priming of eigencentrality in the allocation of community splits could be desirable for particular applications, e.g., when analyzing networks with highly heterogeneous eigencentrality across the nodes.  This will be particularly important in networks whose node eigencentrality is not fully captured by the degree centrality~\cite{Bonacich2007}, e.g.,
when a low-degree individual is connected to high degree others or in which a high-degree node is only connected to low degree others.

Finally, an interesting property of the Ruelle-Bowen random walk is its universality.
Any linear dynamics $x_{t+1}=x_t \, A$, where $x_t$ is a row vector of real entries over the nodes and $A$ is a nonnegative primitive matrix, can be transformed to make it interpretable as a random walk~\cite{ES:81}. 
Hence, besides consensus, heat diffusion, linearized synchronization, etc, random walks can also be used to represent a wider class of dynamics on networks.

\section{Markov Stability for Directed Graphs}

Another advantage of the dynamical framework for community detection introduced above is that it extends naturally to directed graphs, whereas the extension of structural quality functions, such as modularity, to the case of directed graphs is not trivial. 
For instance, although it has been argued~\cite{leicht2008community,Arenas} that the null configuration model in modularity should become $d_\text{in} d^T_\text{out}/2m$ in order to account for the directionality of the links, this choice and justification of the null model for directed graphs is not unique. Under our dynamical viewpoint,  the notion of community becomes that of flow community, and the relevant centrality is pagerank with its associated null model, as we show below.

Consider the simple random walk for a directed graph wit the (non-symmetric) adjacency matrix:  $A\neq A^T$. Each node has an in-degree, collected in the vector $d_{in}=A^T\1$, and an out-degree, collected in the vector $d_{out}=A\mathbf{1}$, i.e., the sum of the weights of the edges directed at and departing from the node, respectively.  The simple random walk in this case is given by 
\begin{align}
\label{eq:RWdir}
\mathbf{p}_{t+1} =  \mathbf{p}_t \, D_\text{out}^{-1}A   = \mathbf{p}_t \,M_\text{dir}
\end{align}
where $D_\text{out} = \mathrm{diag}(d_\text{out})$ and $M_\text{dir}=D_\text{out}^{-1}A$. For nodes where $d_{\text{out},i}=0$, we set $D_\text{out}(i,i)=1$.

For simplicity, consider first the case when the graph is strongly connected and aperiodic. Then the random walk~\eqref{eq:RWdir} is ergodic and
has a unique, stationary distribution $\pi_\text{dir}$ corresponding to the dominant left eigenvector of $M_\text{dir}$. The stationary distribution $\pi_\text{dir}$ is called pagerank,  a key measure of centrality in directed graphs~\cite{perra2008spectral}. 
We can then define the directed Markov Stability based on the random walk~\eqref{eq:RWdir}, which has the same form as~\eqref{eq:stability_disc}~and~\eqref{eq:autocovariance}.  This quality function can be used the same way as the undirected version to extract multiscale structure in graphs by using the Markov time $t$ as a resolution parameter.
The directed Markov Stability at time $t=1$ which, following~(\ref{eq:modul_stab}) above, corresponds to our quality function most closely related to `directed modularity' : 
\begin{equation}
\label{eq:dir_modul}
r_{\text{dir}, 1}=\trace \left[ H^T (\Pi_\text{dir} D_\text{out}^{-1}A- \pi_\text{dir}^T \pi_\text{dir}) H \right].
\end{equation}
Note that the null model we obtained here corresponds to the outer product of the normalized pagerank vector  $\pi_\text{dir}^T \pi_\text{dir}$, in lieu of in- and/or out-degree vectors~\cite{leicht2008community,Arenas}.  

\begin{figure}[!ht]
\begin{center}
\includegraphics[width=0.4\textwidth]{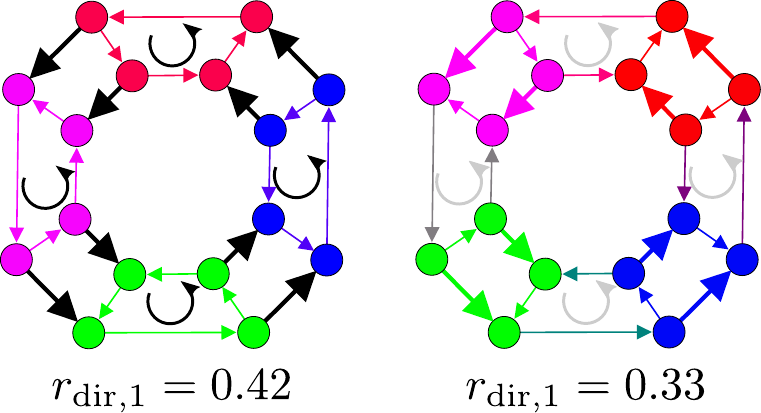}
\caption{{\bf Directed Markov Stability versus extensions of modularity.}
\small{In this toy network~\cite{rosvall}, the weight of the bold links is twice the weight of the other links. The partition on the left (indicated by different colors) optimizes directed Markov Stability~\eqref{eq:dir_modul}, which intrinsically contains the pagerank as a null model. The partition on the right instead optimizes an extension of modularity based on in- and out-degrees~\cite{leicht2008community,Arenas}. Hence directed Markov Stability produces flow communities, whereas the extension of modularity ignores the effect of flows.}}
\label{fig:directed}
\end{center}
\end{figure}

Clearly, using~\eqref{eq:dir_modul} gives different results to structural versions of directed modularity based on in- and out-degree null models. 
While optimization of \eqref{eq:dir_modul} favours partitions with persistent flows of probability within modules, modularity favours partitions with high densities of links and is blind to the flow actually taking place on these links. To illustrate the difference, consider the toy example given by~\cite{rosvall} (Fig.~\ref{fig:directed}), on which the directed random walk is ergodic. In this case, optimizing the in/out-degree modularity of this toy network leads to a partition where heavily weighted links are concentrated inside communities, as expected. On the other hand,  optimization of directed Markov Stability leads to a partition where flows are trapped within modules. It is also interesting to stress that the partition that optimizes~\eqref{eq:dir_modul} also optimizes the map equation proposed by Rosvall and Bergstrom\cite{rosvall}. For an independent study of directed modularity based on other arguments, see Kim et al~\cite{fKim}. 

Our definition of directed Markov Stability relies on the condition that the dynamics is ergodic.
When the directed network is not ergodic, it is common to generalize the standard random walk by incorporating a random teleportation term (also known as `Google teleportation'). If the walker is located on a node with at least one outlink, it follows one of those outlinks with probability $\tau \in (0,1)$. Otherwise, with probability $1-\tau$, the random walker teleports with a uniform probability to a random node. 
Instead of $M_\text{dir}$, the new transition matrix of the random walk~\eqref{eq:RWdir} becomes:
\begin{equation}
 M_\text{dir}(\tau) = \tau M_\text{dir} + \left[(1-\tau) I + \tau \,
    \mathrm{diag}(a)\right]\frac{\mathbf{1}\mathbf{1}^T}{N},
  \label{eq:teleport}
\end{equation}
where the $N \times 1$ vector $a$ is an indicator for dangling nodes: 
$a_i=1$ if $d_{\text{out},i}=0$ (and the corresponding row of $M_\text{dir}$ is assumed to be zero) and $a_i=0$ otherwise. Upon visiting a dangling node, a random walker is teleported with
 probability 1.
It is customary to use the value $\tau=0.85$.  
The teleportation scheme is known to make the dynamics ergodic and to ensure the existence of a single stationary solution $\pi_\text{dir}(\tau)$ that is an attractor of the dynamics. Indeed, teleportation is sometimes introduced even in the ergodic case to improve the numerical convergence of pagerank computation. 

Finally, we remark that, as for the undirected case, there are continuous-time versions of directed Markov Stability. The simplest is given by the corresponding Kolmogorov equation:
\begin{equation}
\label{eq:norm_dir}
\frac{d\mathbf{p}}{dt} = - \mathbf{p}\;[I - M_\text{dir}(\tau)], 
\end{equation}
and our discussion above applies to these processes too. An application to a large graph of airport connections is presented in the next section.  See also~\cite{Beguerisse2014riots} for an application to social network analysis.

\section{Computational methodology and practical considerations}

Given a network, and based on modelling considerations or other assumptions, we can choose a discrete- or continuous-time Markov process to scan dynamically the structure of the graph at all scales. As shown in the toy example of Figure~\ref{fig2}, the optimization of the chosen Markov Stability across time leads to a sequence of partitions that are optimal at different time scales.  The extraction of these optimized partitions is the first step to uncover the multi-scale modular structure of the network (if present), but the practical application of the method still involves at least two non-trivial steps, which we now discuss in conjunction with several larger examples. Although the examples in this section exhibit a relatively hierarchical community structure, in Supp.Inf. we illustrate and measure quantitatively non-hierarchical multi-scale structures.

\label{sec:methods}
\subsection{Optimization of Markov Stability}
Although it has been shown that modularity optimization is NP-hard~\cite{NP}, several heuristic algorithms have been proposed to provide satisfactory solutions, in the sense that they efficiently recover planted solutions in benchmark graphs, or that they can uncover groups that are clearly meaningful (e.g. classes in a school social network, etc)~\cite{fortunato2010community}. 
It has also been shown that in real-world examples the modularity landscape over partitions tends to exhibit large rugged plateaux, making it possible to find an approximately optimal partition~\cite{Good}.

We will now show that it is always possible to rewrite the Markov Stability for any choice of random walk as the modularity of another symmetric graph. This observation has important practical implications, as it makes it possible to use any modularity-maximization algorithm, e.g. spectral or greedy, for the optimization of any version of Markov Stability.
For example, consider the discrete-time stability $r(t)=  \trace \left[ H^T(\Pi M^t - \pi^T \pi)H \right]$, for transition matrix $M$ and the corresponding centrality $\pi$. It is easy to see that this is the usual modularity for the graph of weighted adjacency matrix $A= (\Pi M^t+(\Pi M^t)^T)/2$, a symmetric matrix of degree sequence $A \1 =\pi^T$. A similar observation holds in continuous time (where the exponential can be evaluated by Pad\'e approximations), and also for the linearized versions of Markov Stability.

Any modularity maximization algorithm can therefore be used for Markov Stability optimization. As some of those algorithms~\cite{Blondel} are empirically known to run in  $O(m \log m)$ on $m$-edge graphs, the most expensive step turns out to be matrix multiplication or computation of the exponential, which limits the application of full Markov Stability to graphs with $N \sim 20000$ nodes. These overheard costs are spared when using the linearized version of Stability, which becomes the most suitable for the multi-scale analysis of very large networks $N > 10^5$.
In our applications below, we have used mainly the Louvain algorithm~\cite{Blondel} adapted to the optimization of Markov Stability\footnote{An efficient code, also with a Matlab interface, can be downloaded at \url{http://wwwf.imperial.ac.uk/~mpbara/Partition_Stability/} or \url{http://michaelschaub.github.io/PartitionStability/}}, although spectral bisection methods~\cite{shimalik} for the generation of optimized partitions yields good results~\cite{JC}.

\begin{figure}[!ht]
\begin{center}
\includegraphics[width=.45\textwidth]{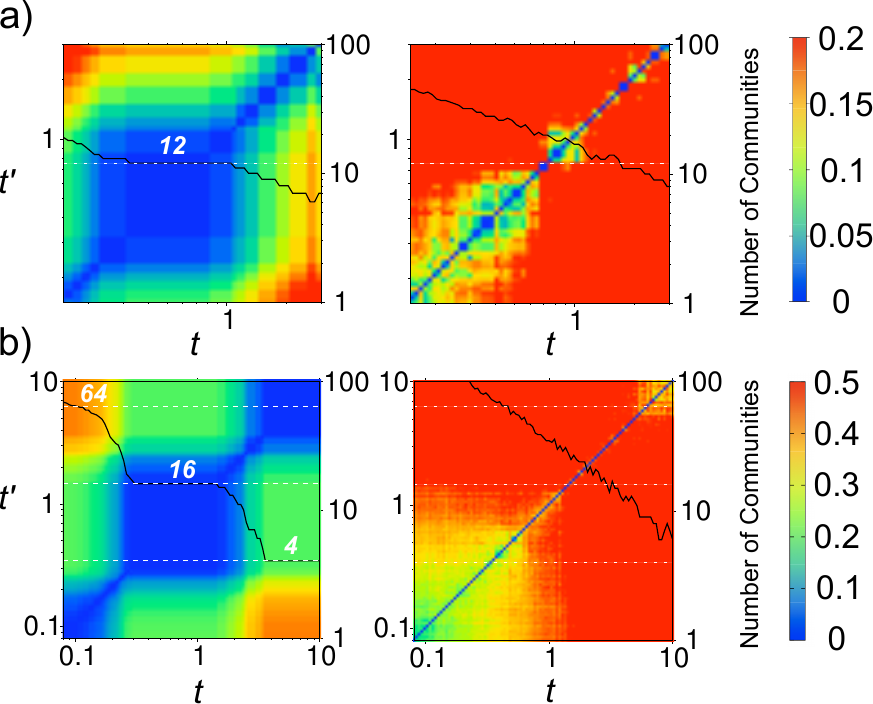}
\end{center}
\caption{{\bf Selecting robust partitions in the sequence of optimized partitions across Markov time.} 
\small{(a) The American football network~\cite{GN} composed of $N=115$ teams is known to be organized into 12 divisions. (Left) The block structure of the normalized variation of information~\eqref{eq:VI} between the optimized partitions at time $t$ and $t'$ 
and a long plateau in the number of communities indicates that the most persistent partition is made of 12 communities, as expected. (Right) The randomized version of the network, where links have been reshuffled while preserving the node degrees, does not exhibit robust communities.
(b) A benchmark hierarchical random network consisting of $N=640$ nodes with $3$ levels: $64$ modules of $10$ nodes; $16$ modules of $40$ nodes; $4$ modules of $160$ nodes~\cite{sales}. 
We use one realization of the benchmark. Similarly to (a), the long plateaux in the number of communities and the block structure with low values of $\hat{V}(\mathcal{P}(t),\mathcal{P}(t'))$ reveal the three levels of the
hierarchy (left). No significant community structure is detected in the randomized network (right).
Both sequences of partitions were obtained optimizing $r_{\text{norm}}(t; H)$ with the Louvain algorithm~\cite{Blondel}.
}}
\label{fig:football}
\end{figure}

\begin{figure}[htp!]
\begin{center}
\includegraphics[width=.45\textwidth]{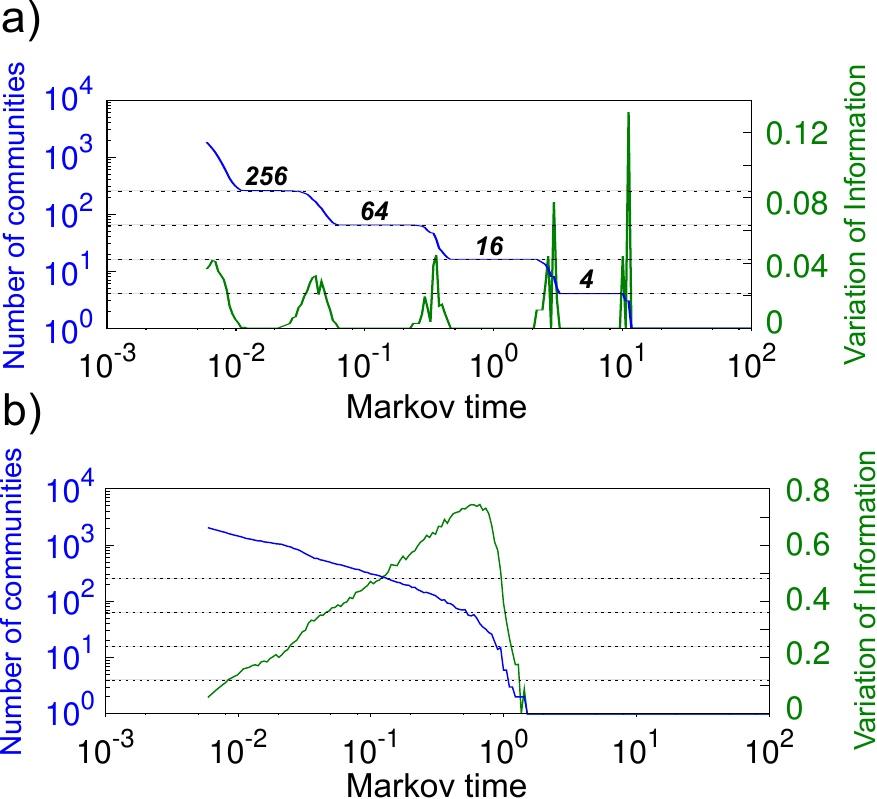}
\end{center}
\caption{{\bf Hierarchical benchmark and statistical tests.} 
\small{Benchmark random network with $N=2560$ nodes and $4$ hierarchical levels 
(with modules of $10$, $40$, $160$ and $640$ nodes)~\cite{sales}. 
(a) The long plateaux in the number of communities (blue) and the dips in the normalized variation of information across time $\hat{V}(\mathcal{P}(t),\mathcal{P}(\lambda t))$ (green)
signal that the four levels of the hierarchy have been detected.  
(b) Same as (a) for a randomized version of the network preserving node degrees: no community structure is found at any scale. In both cases, $\lambda=20/19$ and the sequences of partitions were obtained optimizing $r_{\text{norm}}(t; H)$ with the Louvain algorithm.}
}
\label{fig:hier}
\end{figure}

\subsection{Robustness of partitions}

Once the sequence of optimized partitions is obtained, we need to select the most relevant scales (partitions) for our description. This is a well-known challenge for multi-resolution methods.
Notions of robustness are usually considered when dealing with NP-hard optimizations to reflect the ruggedness of the landscape of the quality function to be optimized~\cite{Good}. 
In our approach, we establish the significance of a particular partition based on its robustness in three different ways~\cite{ronhovde, Delmotte2011, stability1, karrer, reichardt,multiscalelambiotte}:  
(i) robust (persistent) across time; 
(ii) robust to small perturbations to the graph; 
and (iii) robust to the optimization algorithm and the starting point of the optimization.
We now exemplify (i) and (iii).  

The basic notion is to evaluate the effect of these perturbing factors on the optimized partition: a partition is robust if such perturbations have little effect on the outcome and the perturbed result remains close to the unperturbed one.
A popular way to compare two partitions $\mathcal{P}_1$ and $\mathcal{P}_2$ is the normalized variation of information~\cite{meila} 
\begin{equation}
\label{eq:VI}
\hat{V}(\mathcal{P}_1,\mathcal{P}_2) = \frac{H(\mathcal{P}_1|\mathcal{P}_2)+H(\mathcal{P}_2|\mathcal{P}_1)}{\log N},
\end{equation}
where $H(\mathcal{P}_1|\mathcal{P}_2)$ is the conditional entropy of the partition $\mathcal{P}_1$ given $\mathcal{P}_2$, i.e., the additional information needed to describe $\mathcal{P}_1$ once $\mathcal{P}_2$ is known assuming a uniform probability on the nodes. The conditional entropy is defined in the standard way for the joint distribution $P(C_1,C_2)$ that a node belongs to a community $C_1$ of $\mathcal{P}_1$ and to a community $C_2$ of $\mathcal{P}_2$. The normalized variation of information $\hat{V}(\mathcal{P}_1,\mathcal{P}_2) \in [0,1]$ has been shown to be a true metric on the space of partitions and vanishes only when the two partitions are identical.

Within the Markov Stability framework, we use this metric to evaluate the persistence of partitions across time. 
By looking for block-diagonal regions with low values of $\hat{V}(\mathcal{P}(t),\mathcal{P}(t'))$, 
as well as plateaux in the number of communities as a function of time~\cite{mason}, 
we can detect the relevant partitions and scales without assuming them \textit{a priori}.    
Two examples of this approach are shown in Fig.~\ref{fig:football}, where we illustrate the detection of the relevant scale (12 communities) in a small real-life network of American football teams ($N=115$), as well as three scales in a hierarchical benchmark random network with $N=640$ nodes. 
The same notion is evaluated in Fig.~\ref{fig:hier}, where we detect 4 hierarchical levels 
in a larger benchmark network with $N=2560$ nodes 
by comparing partitions across time using the scaling factor $\lambda$ 
to evaluate $\hat{V}(\mathcal{P}(t),\mathcal{P}(\lambda t))$.

\begin{figure}[t]
\begin{center}
\includegraphics[width=.45\textwidth]{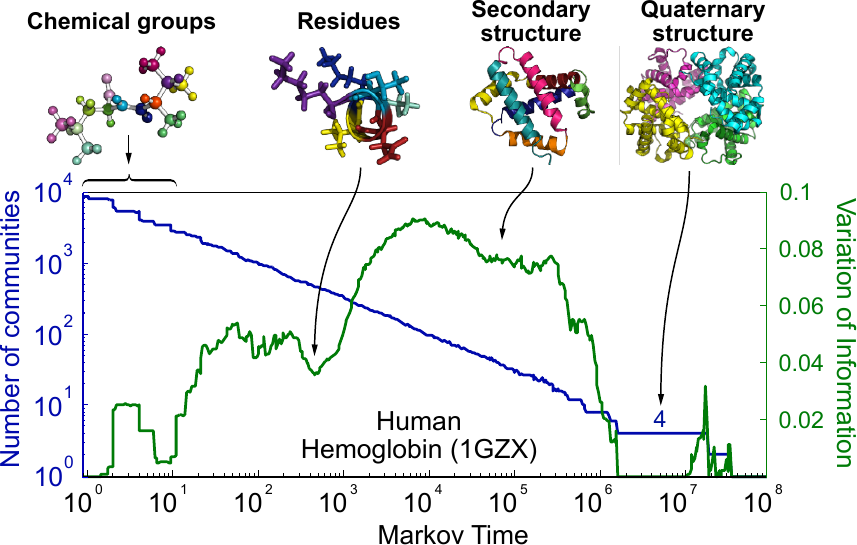}
\end{center}
\caption{{\bf Finding robust communities at multiple scales in the atomic network of
hemoglobin, a protein tetramer.} 
\small{The atomic network of the protein is generated as detailed in Refs.~\cite{Delmotte2011, Delmotte_thesis} using physico-chemical potentials and atomic X-ray crystallographic data (PDB file: 1GZX). This weighted, undirected network has $N= 8757$ nodes (atoms) and $12813$ edges (bonds). The multi-scale nature of our method reveals relevant communities across scales, from small chemical groupings to large-scale conformations, signalled by dips of the normalized variation of information. These dips deviate significantly from chemically-consistent randomized versions of the network (not shown; see~\cite{Delmotte2011, Delmotte_thesis}).  Note the long plateau and dip of the normalized variation of information for the 4-way partition, corresponding to the identification of the four monomers in the hemoglobin tetramer. Here the combinatorial version of Markov Stability $r_{\text{comb}}(t; H)$ was optimized, as it is more closely matched to the vibrational dynamics of the protein network.}
}
\label{fig:protein}
\end{figure}

\begin{figure*}[!ht]
\begin{center}
\includegraphics{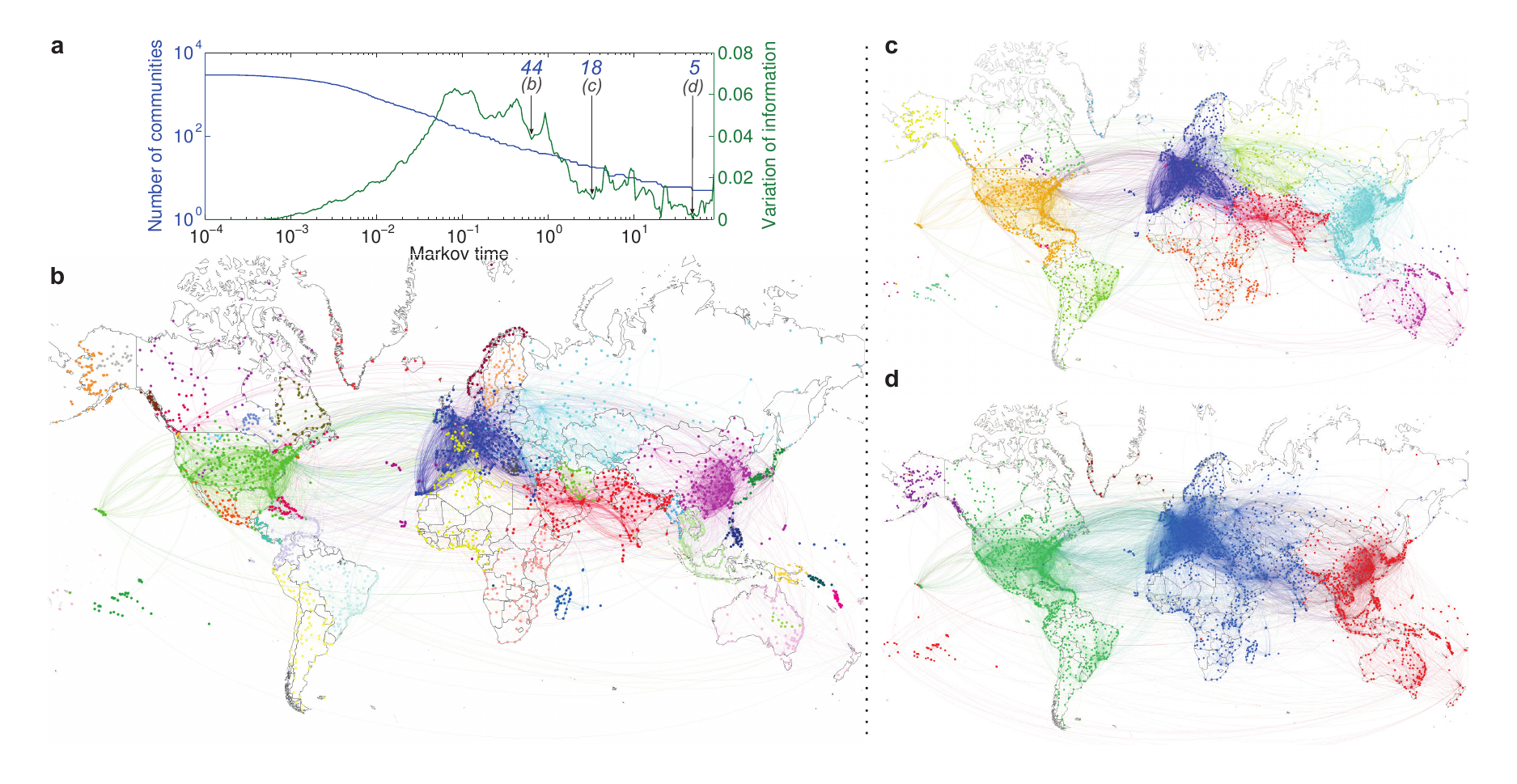}
\end{center}
\caption{{\bf Flow communities at multiple scales in an airport network.} 
\small{The airport network~\cite{airports2011} contains $N=2905$ nodes (airports) and $30442$ weighted directed edges. 
The weights record the number of flights between airports (i.e., the network does not take into account passenger numbers, just the number of connections). Representative partitions at different levels of resolution with (b) 44, (c) 18 and (d) 5 communities are presented. The partitions correspond to dips in the normalized variation of information in (a) and show persistence across time (see Suppl. Info.).}}
\label{fig:airport}
\end{figure*}

In addition to the robustness of partitions based on persistence across time, it is also helpful to evaluate the robustness of the solution with respect to the optimization. 
We do this by repeating the Louvain optimization many times 
(in excess of 100 random initial seeds for each Markov time) and evaluating the
average normalized variation of information within the \textit{ensemble} of optimized solutions. If a partition is robust to the optimization, we expect a small value (or a dip) in the normalized variation of information of the ensemble of optimized solutions, signaling a relevant partition. This robustness to the optimization probes the ruggedness of the landscape and can be tested for different optimization algorithms~\cite{Good}. Here we use
the Louvain algorithmic heuristic, which has been shown to perform well both in benchmarks and real-life examples~\cite{multiscalelambiotte}.
In Figures~\ref{fig:protein}~and~\ref{fig:airport}, we show the application of this approach to two large networks: an undirected, weighted atomic protein network with $N=8757$ nodes; and a directed, weighted network of airport connections with $N=2905$ nodes. 
In both networks, we find relevant structure at different resolutions. Of note is that our results in the protein network are able to identify partitions corresponding to relevant chemical structures (involving only a few nodes), through secondary structures such as helices (involving several hundreds of atoms) to large conformational domains and, importantly, the subunits (involving several thousands of atoms).  In the case of the airport network, the different levels of resolution reveal geographical and socio-political groupings. In this case, the directed character of Markov Stability is able to reveal communities with specific  flow characteristics, including regions with focalized entry points coupled to a local asymmetric distribution network (e.g., Alaska and Greenland). 

The selection of the relevant scales is still an open area of research in multiscale community methods and has strong links with non-convex optimization. Our notions of robustness reveal that the optimized partitions found at peaks of the variation of information tend to be hybrid combinations of natural partitions with non-uniform resolution, splitting some but not all the coarser communities, thereby explaining a high sensitivity to the random seed or Markov time. In other cases, such peaks correspond to the coexistence of a few `good' partitions, which might indicate a tendency to flip between such outcomes and, hence, a lack of robustness. In this sense, the peaks in the variation of information tend to signal the separation between the relevant scales in the community structure of the network, and can also be related to the existence of non-hierarchical (yet multi-scale) community structure (see Supp.\ Info.\ for some examples). These topics will be the object of further work.


\section{Discussion}

Our work emphasizes the importance of choosing proper dynamical processes in order to uncover information in networked systems. Here, we have focused on random walk processes, which are known to be mathematically equivalent to a broad range of diffusive processes: heat diffusion, evolution on a (free) energy landscape~\cite{wales2003energy}, opinion dynamics on social networks and other kinds of consensus problems~\cite{fax2004information,ROS-JAF-RMM:07}, linearization of synchronization~\cite{StevenH20001,syn2} and power networks~\cite{dorfler2013synchronization}, among others. Importantly, using the random walk corresponding to the natural dynamics of the system allows us to find its central nodes (according to its intrinsic centrality measure) and to recover dynamically meaningful communities, i.e., the communities of nodes that best retain the diffusive flow for a certain time scale. If there is no intrinsic dynamics in the system, and hence no unique choice for the exploratory Markov dynamics, our approach provides tools to understand the effect of the different choices of random walks and associated centrality measures on the community structure obtained through Markov Stability optimization. 

\begin{figure*}
\begin{center}
\begin{tabular}{|c|c c c c|}
\hline 
   & Simple Random Walk            & Normalized Laplacian& Combinatorial Laplacian &  Ruelle-Bowen \\
   \hline 
   Type &   Discrete-time       & Continuous-time          & Continuous-time       & Discrete-time  \\

    Node centrality &   Degree       & Degree          & Uniform       & Eigencentrality  \\
     Linearized Stability &  Potts model\cite{ReichardtBornholdt06}          & Potts model\cite{ReichardtBornholdt06}  &  Potts model\cite{traag2011narrow,ReichardtBornholdt06}  &             \\
     Time-one (linearized) stability & Modularity\cite{NG}  &  Modularity\cite{NG}  &   Modularity\cite{NG}  &					\\
     Null model &      Configuration model  &    Configuration model     & Erdös-Rényi   &            \\
     Spectral Algorithm &    Shi-Malik\cite{shimalik}      & Shi-Malik\cite{shimalik}       &  Fiedler\cite{Fiedler73,Fiedler75}           &   Sussman\cite{sussman2012consistent}       \\
     \hline 
\end{tabular}
\end{center}
\caption{{\bf Summary of the dynamics-based Markov Stability framework and connections
with centrality measures, and other clustering and community detection methods in the literature.} 
}
\label{tab:summary}
\end{figure*}

More generally, our approach provides a unified viewpoint for a number of existing approaches, as summarized on Fig.~\ref{tab:summary}, and 
Our approach paves the way for the development of metrics and algorithms that exploit real-world non-Markovian random walks~\cite{Rosval} or incorporate non-trivial temporal patterns into diffusive models~\cite{Hoffman}. 
Our work also opens perspectives in community detection by providing a dynamical interpretation of quality functions, and by interpreting the standard null-model paradigm in terms of stationary distributions~\cite{NG,GN}. The dynamical approach that we advocate here, not only generalizes the null model paradigm, but can also lead to fundamentally different quality functions. For instance, even the simple random walk on a directed graph leads to a Stability function containing the pagerank, which is not expressible in terms of combinatorial quantities, hence different from any null-model-based variant of modularity. The dynamic and null-model paradigms do overlap in a number of interesting cases. We have shown that for undirected networks, the two most common continuous-time dynamics, described by the normalized and combinatorial Laplacians, correspond to the two most meaningful null models, i.e., the configuration model and the Erdös-Rényi model. Through the intuition gained from the corresponding dynamics, we reinterpret the Erd\"os-R\'enyi null model (long considered as inferior in the null-model literature~\cite{NG,GN}) and show that it is linked to an optimization that tends to produce node-balanced communities, and can be more relevant under particular dynamical processes, consistent with the findings of Traag et al~\cite{traag2011narrow}. The exploration of alternative random walks, such as the Ruelle-Bowen walk, also highlights the capability of introducing alternative measures of centrality and extending community detection to include non-standard Markov processes.

\section*{Acknowledgments}  
We thank M.T.\ Schaub and A.\ Delmotte for extended discussions and for help with some of the figures and computations. We also acknowledge T.\ Evans, S.N.\ Yaliraki, M.\ Draief, H.J.\ Jensen, V.\ Blondel, M.A.\ Porter, V.\ Latora, M.\ Rosvall and J.\ Saram\"aki for fruitful discussions.  J.-C.D.  acknowledges support from the Belgian Programme of Interuniversity Attraction Poles, an Action de Recherche Concert\'ee (ARC) of the French Community of Belgium, and  FP7 STREP project EULER of the European Commission. 
R.L. acknowledges support from FNRS and project Optimizr of the European Commission. M.B. acknowledges funding from grant EP/I017267/1 of the UK EPSRC (Engineering and Physical Sciences Research Council) under the \textit{Mathematics Underpinning the Digital Economy} program.
 
\bibliography{biblio}
\bibliographystyle{IEEEtran}

\appendix

\subsection*{Hierarchical versus multiscale organization}

\begin{figure}[!ht]
\begin{center}
\includegraphics[width=0.48\textwidth]{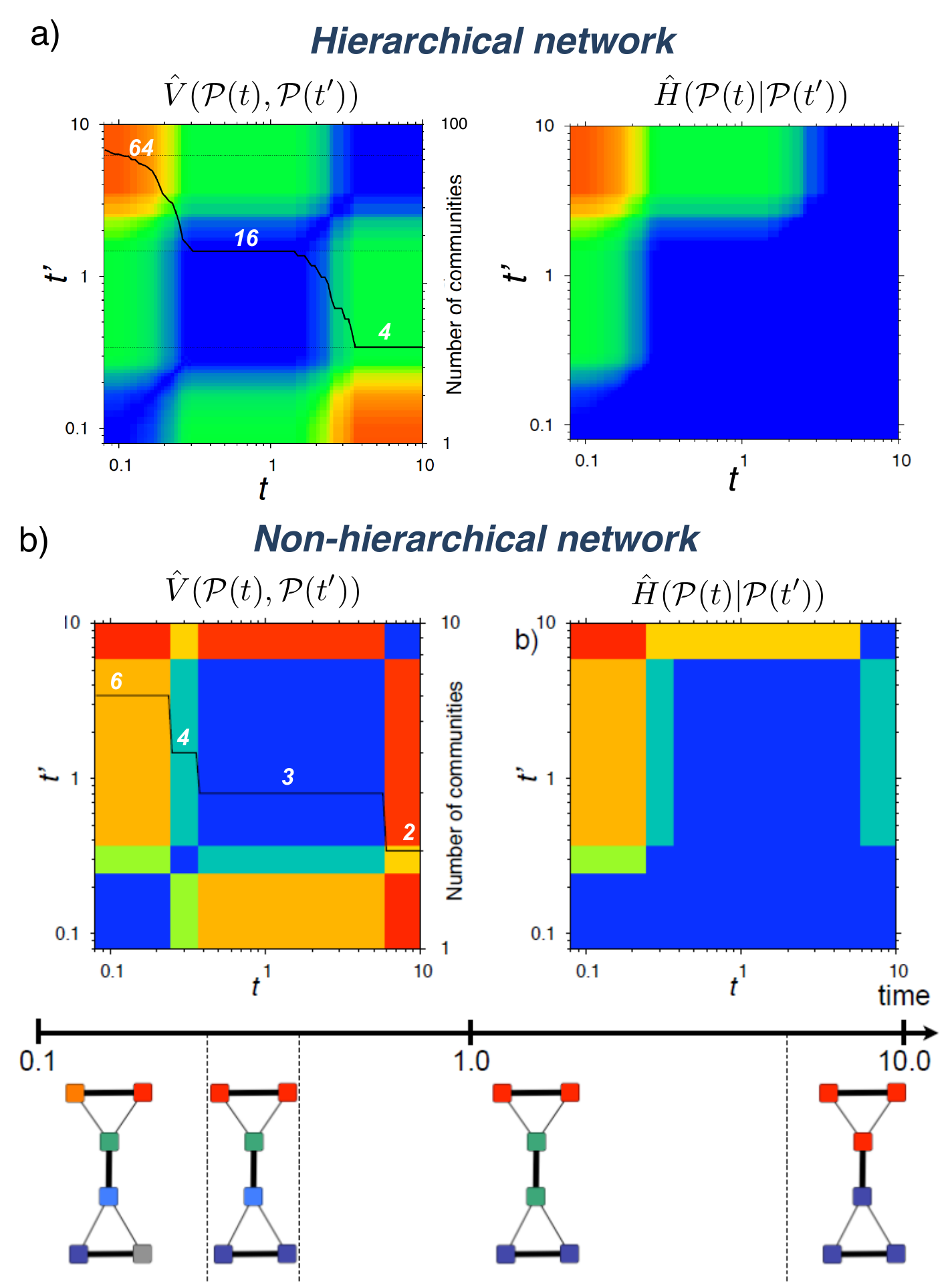}
\end{center}
\caption{{\bf Lack of hierarchy in a toy network.} 
We optimize the Markov Stability $r_\text{norm}(t)$ of: (a) the hierarchical model in Fig.~\ref{fig:football}b, and (b) a toy network (bottom panel) with 6 nodes and links of different strength (thick links of weight 5, thin links of weight 1). 
(a) The normalized variation of information $\hat{V}(\mathcal{P}(t),\mathcal{P}(t'))$ (left, same as in Fig.~5b), indicates the presence of three levels of a hierarchy. The conditional entropy 
$\hat{H}(\mathcal{P}(t)|\mathcal{P}(t'))$ (right) reveals that the obtained community structure respects a strict hierarchy, although the Louvain optimization method does not impose such a hierarchical structure 
\textit{a priori}.
(b) For the toy network, the normalized variation of information $\hat{V}(\mathcal{P}(t),\mathcal{P}(t'))$ and the number of communities (left) reveal a sequence of partitions with 6, 4, 3, and 2 communities (shown bottom). The 3-way partition is especially robust. In this case, however, the sequence of uncovered partitions is not hierarchical since the three-way partition is not nested into the two-way partition. This is revealed by the conditional entropy $\hat{H}(\mathcal{P}(t)|\mathcal{P}(t'))$ (right): there is a region of $t>t'$ in which 
$\hat{H}(\mathcal{P}(t)|\mathcal{P}(t'))>0$.
}
\label{fig:nonhier}
\end{figure}

Our use of time as a resolution parameter enables Markov Stability to detect robust partitions at different scales without imposing \textit{a priori} the coarseness of the partitions. Although some of the methods used to optimize Markov Stability can lead to hierarchical community structure (e.g., the use of recursive bipartitions via Shi-Malik~\cite{JC}), we also use optimization heuristics that do not impose such a constraint (e.g., the use of the Louvain algorithm~\cite{Blondel} optimized independently at each time). It is then interesting to check whether or not the sequence of partitions is compatible with a hierarchical organization. This problem requires the introduction of a quantity that measures whether the communities at time $t'$ are nested into the communities at a subsequent time $t>t'$. A well-known information theoretic measure that is particular adapted for such a purpose is the normalized conditional entropy: 
\begin{equation}
\label{eq:conditional_entropy}
\hat{H}(\mathcal{P}(t) |\mathcal{P} (t'))=\frac{H\left(\mathcal{P}(t)|\mathcal{P}(t')\right)}{\log N},
\end{equation}
which is also constrained to the interval $[0,1]$ but is now an asymmetric quantity that vanishes only if each community of $\mathcal{P}_t$ is the union of  communities of $\mathcal{P}_{t'}$. The combined knowledge of  $\hat{V}$ and $\hat{H}$ therefore allows us to uncover the significant partitions of the system and to verify if those partitions are organized in a hierarchical manner.
For instance, the benchmark in Fig.~\ref{fig:football}b is clearly hierarchical, as can be seen in Figure~\ref{fig:nonhier}a, whereas the toy network in Fig.~\ref{fig:nonhier}b shows that the sequence of the optimal partitions is not necessarily hierarchical.

\subsection*{Consistency of robustness measures in the airport network}
\begin{figure*}[!ht]
\begin{center}
\includegraphics[width=0.95\textwidth]{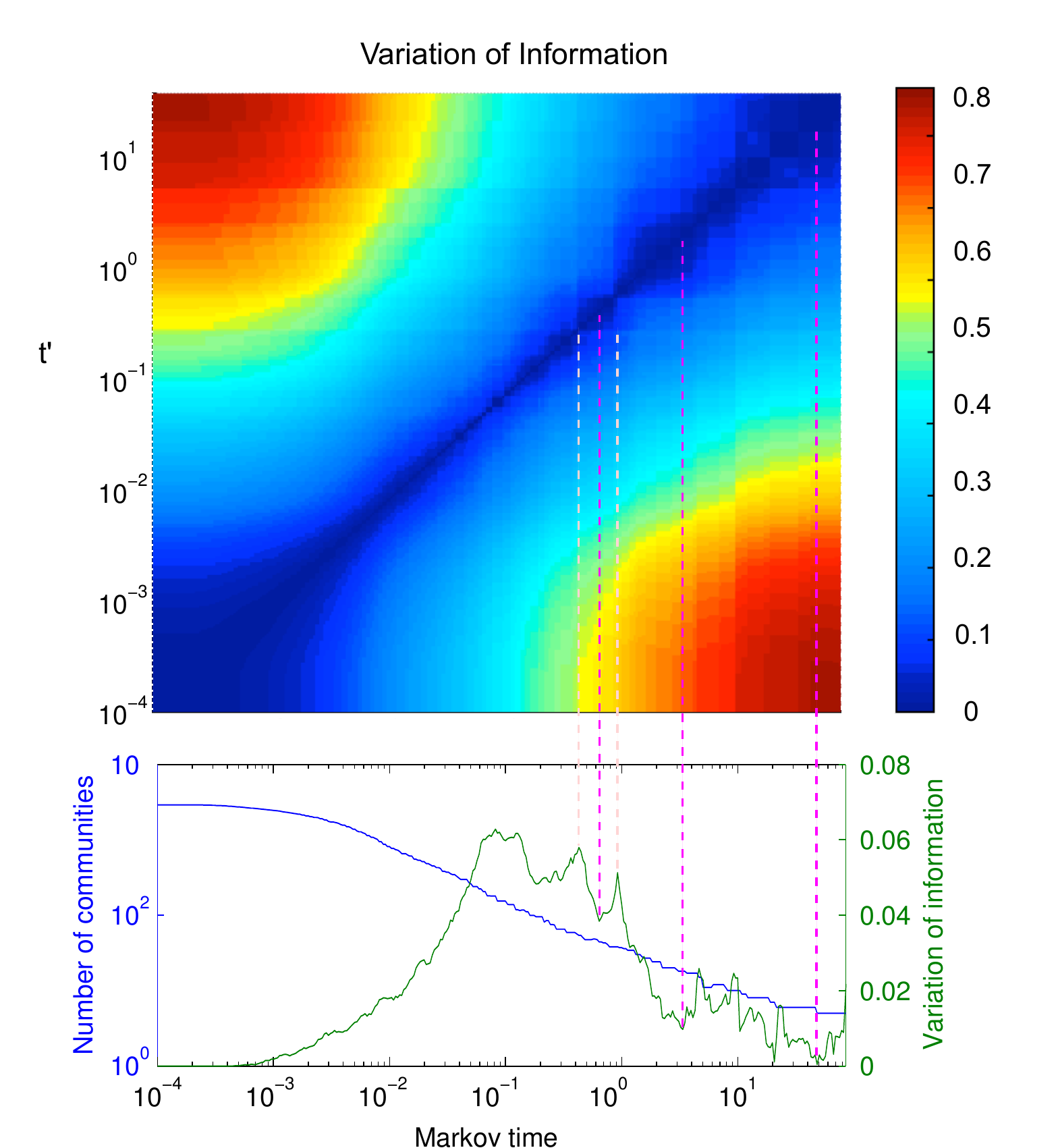}
\end{center}
\caption{{\bf Consistency of two robustness measures: persistence across time and optimization.}
This figure complements Fig.~\ref{fig:airport}.
(Top) Colormap of the normalized variation of information $\hat{V}(\mathcal{P}(t),\mathcal{P}(t'))$ for the optimized partitions of the airport network across time.  The dark blue blocks indicate 
plateaux of similar partitions (see Fig.~\ref{fig:football} and Figure~\ref{fig:nonhier}a). 
(Bottom) The normalized variation of information of the ensemble of solutions with respect to the random seed of the optimization (same as in Fig.~\ref{fig:airport}a).
The Markov times delimiting the blocks (top) correspond to peaks of the normalized variation of information of the ensemble of solutions (bottom), while the dips fall within the squares. Some of these dips have been presented as representative partitions in Fig.~\ref{fig:airport}.}
\label{fig:nvi_airp}
\end{figure*}

As a complement to Fig.~\ref{fig:airport}, Fig.~\ref{fig:nvi_airp} shows that the dips in the normalized variation of information of the ensemble of solutions (presented in Fig.~\ref{fig:airport}) are consistent with the presence of block-structure in the normalized variation of information between the optimized solutions found across time $\hat{V}(\mathcal{P}(t),\mathcal{P}(t'))$.

\end{document}